\documentclass[a4paper]{article}
\usepackage{graphicx}

\title{Self-dual random-plaquette gauge model\\and the quantum toric code}
\author{Koujin Takeda\footnote{E-mail address:
takeda@stat.phys.titech.ac.jp} ~and Hidetoshi Nishimori\\
Department of Physics, Tokyo Institute of Technology,
\\
Oh-okayama, Meguro, Tokyo 152-8551, Japan}
\date{}
\begin{document}

\maketitle
\vspace{1.5cm}
\begin{center}
{\bf Abstract}
\end{center}

We study the four-dimensional $Z_2$ random-plaquette
lattice gauge theory as a model of topological quantum memory, the toric code in
 particular.
In this model,
 the procedure of quantum error correction 
 works properly in the ordered (Higgs) phase, and 
 phase boundary between the ordered (Higgs) and disordered (confinement) phases gives
the accuracy threshold of error correction. 
Using self-duality of the model in conjunction with the replica method,
we show that this model has exactly the same mathematical structure
as that of the two-dimensional random-bond Ising model,
which has been studied very extensively.
This observation enables us to derive a conjecture on the exact location of
the multicritical point (accuracy threshold) of the model,
 $p_c=0.889972\ldots$,
and leads to several nontrivial results including bounds on the
accuracy threshold in three dimensions.

\vspace{1cm}
\clearpage

 \section{Introduction}
 The lattice gauge theory has been studied mainly from the standpoint of
 particle physics.
 Many interesting results have been known for years for systems with
 spatially uniform couplings \cite{lattice}.
 The lattice gauge theory with randomness in couplings has not
 been a target of active studies mainly due to lack of physical motivation.
 However, it was recently proposed that the random lattice gauge model can be
 related closely with error-correcting processes of
 the toric code, an interesting example of quantum information storage
 \cite{Kitaev,DKLP,WHP}.
 This observation serves as a strong motivation to investigate
 the random case systematically.

 In Refs. \cite{Kitaev,DKLP,WHP}, the two-dimensional ($2D$)
 $\pm J$ random-bond Ising model (RBIM)
 and the $3D$ $Z_2$ random-plaquette gauge model (RPGM) were studied in the context
 error corrections of the toric code.
 More precisely, the error-correcting processes of the bond-qubit model,
 where a qubit resides on each 
 bond of the lattice, can be mapped to the RBIM.
 The plaquette-qubit model, where a qubit is located
 on each plaquette, is similarly related to the RPGM
 as far as error-corrections are concerned.
 
 In the toric code, we assume that the phase error and
 bit flip error may occur on each qubit due to decoherence and
 interactions with the environment.
 Fortunately, if the number of errors is not very large,
 we can determine the ``error chains'' or ``error sheets''
 from the measured syndrome, or the positions
 where check operators give nontrivial signs ($-1$).
 We can then correct the errors by applying appropriate operators to the relevant
 error positions,
 and the encoded information, which is the simultaneous eigenstate of all check operators
 with trivial (unit) eigenvalues, is recovered.
 However, if the error rate becomes larger than a threshold
 (accuracy threshold), this procedure of error correction 
 fails because nontrivial ambiguities arise in the error positions,
 and consequently the encoded information is lost. 
 It is therefore very important to obtain the correct value of the accuracy threshold.

 The corresponding statistical models such as the RBIM or the RPGM
 have two parameters: the temperature $T$ and the
 probability $1-p$ for the interaction on a bond (plaquette) to be negative,
 which corresponds to the error rate in the context of toric code.
 The error-correcting properties of the toric code are related to
 the ordering behaviour of the RBIM/RPGM on a line in the $p$-$T$ plane
 (phase diagram).
 Error correction can be performed successfully
 in the magnetically ordered phase in the $p$-$T$ plane of the RBIM (RPGM),
 which means stable storage of the encoded information,
 while in the disordered phase error correction fails and
 the encoded information is lost.
 The transition point (the multicritical point on the Nishimori line \cite{NL})
 is thus supposed to give the accuracy threshold of error correction.
 It is therefore crucial to know the correct location of the multicritical
 point of these models for the design of reliable quantum storage, for example.
 
 These recent results have lead us to the study of the $4D$ $Z_2$ RPGM which
 can be related to the $4D$ toric code.
 A part of the reasons to investigate the $4D$ model, not $2D$ or $3D$ cases,
  is that the system
 with spatially uniform coupling is known to be self dual
 in $4D$ \cite{Wegner}, which
 serves as a powerful tool of analyses.
 In addition, the model of plaquette toric code on the $3D$ cubic
 lattice under repeated measurement is also related to the $4D$ RPGM.
 As has been mentioned above,
 the $4D$ plaquette gauge model has a useful property of self duality,
 and we will analyze the $4D$ RPGM
 using the duality technique.
 The formalism of duality transformation used in this paper is due
 to Wu and Wang \cite{WuWang}.
 They proposed the method of duality transformation 
 using Fourier transformation
 for the analysis of $2D$ non-random multi-component spin models.
 Recently it has been shown in Ref. \cite{MNN} that the Wu-Wang duality is 
 applicable to studying the $2D$ {\em random} spin model.
 We show that the duality analysis developed in Ref. \cite{MNN}
 is also applicable to the study of the $4D$ RPGM.
 Our main conclusion is that the mathematical structure determining
 the multicritical point of the $4D$ RPGM
 between the ordered (Higgs) and the disordered (confinement) phases 
 is closely related to that of the $2D$ RBIM, whose property
 has been studied extensively for years.
 This suggests that the accuracy threshold of the $4D$ plaquette toric code
 coincides with the $2D$ bond toric code.
 We also elucidate the correspondence
 between the $4D$ plaquette qubit toric code and the $4D$ RPGM.
 
 The $3D$ RPGM can be viewed as a model of the $3D$ toric code as well.
 We also study the phase
 structure of the $3D$ RPGM using the Wu-Wang duality and give useful bounds on the
 value of accuracy threshold of error correction.

 This paper is organized as follows. 
 In Sec. 2 we review the toric codes for the $2D$ bond model and the $4D$
 plaquette model.
 We also elucidate the relation between the toric code and the
 random spin systems such as the RBIM and the RPGM. 
 In Sec. 3 we explain the Wu-Wang duality technique using the $2D$ Ising model on
 the square lattice, which is one of the simplest self-dual models.
 In Sec. 4 self-duality of the $4D$ plaquette gauge model without randomness is explained. 
 In Sec. 5 we incorporate randomness to the $4D$ plaquette gauge model and
 analyze it using duality and the method of
 averaged Boltzmann factor in conjunction with the replica method
 proposed in Ref. \cite{MNN}. 
 In Sec. 6 the $3D$ RPGM is investigated using its dual representation.
 In the dual space, the $3D$ RPGM can be transformed to an $n$-replicated spin system
 with spatially uniform couplings.  
 Relation to the $3D$ toric code is also discussed.
 Section 7 is devoted to conclusion and discussions.

 \section{Toric code in 2$D$ and 4$D$}

 For clarity of presentation, let us first review the toric code for
 quantum error correction and its relation to spin and gauge models.
 The toric code is a method to encode quantum information for its stable storage
 using topological nontriviality of the manifold on which qubits are located.

 It is useful to start with the $2D$ toric code
 to explain the basic idea  \cite{Kitaev}.
 In this case the toric code is defined on the $2D$ square lattice
 on the torus, and
 qubits reside on bonds (Fig. 1).
 We represent the two states of the qubit by 2-vector
 ${\scriptsize \left( \begin{array}{c} 1 \\ 0 \end{array} \right)}$ and
 ${\scriptsize \left( \begin{array}{c} 0 \\ 1 \end{array} \right)}$.
 Any state of a specified bond can be expressed as a linear
 combination of these vectors. A quantum state of the whole system is 
 given by the direct product of these states.
 The stored quantum information is chosen as a linear
 combination of such direct products which is a simultaneous eigenstate $|\Psi \rangle$
 of all check operators (to be defined below) with trivial eigenvalue 1.
 
 This system suffers from errors caused by decoherence and interactions
 with the environment, and the errors change the states of qubits on bonds.
 There are two types of errors; one is a bit-flip error which
 flips the qubit to the other state, and the other is a phase error to change
 the phase of qubit. We express these errors
 by the operations of Pauli matrices to the original state.
 Existence of these errors on
 bond $l$ is expressed as $\sigma^x_l  |\Psi \rangle$ (bit-flip error) and
 $\sigma^z_l  |\Psi \rangle$ (phase error). 
\begin{figure}
\begin{center}
  \includegraphics[width=11cm]{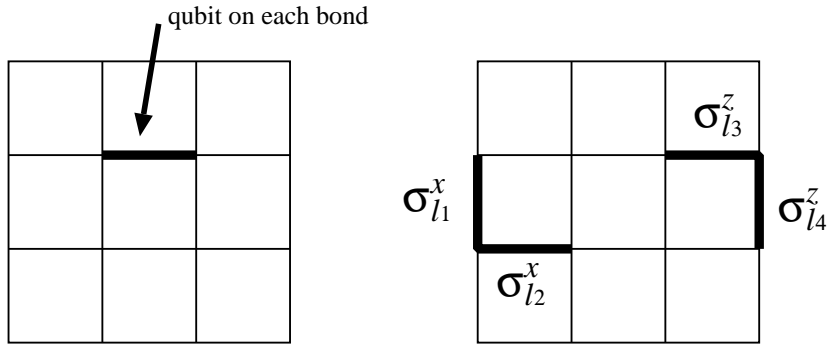}
\label{toriccode1}
\caption{Qubits reside on bonds in the toric code (left). Possible error chains (right).}
\end{center}
\end{figure}
 We can check if these errors have occurred
 by the operation of check operators.
 For this purpose, we define two kinds of check operators,
 \begin{eqnarray}
 X_i & = & \bigotimes_{i \in \partial l} \, \sigma^x_l, \\
 Z_P & = & \bigotimes_{l \in \partial P} \, \sigma^z_l, 
 \end{eqnarray}
 where $X_i$ is defined on each site and $Z_P$
 on each plaquette (shown on the left of Fig. 2).
 All check operators at any sites or plaquettes commute with each other,
 $[X_i, Z_P]=0$,
 and hence simultaneous eigenstates of these operators exist. The
 eigenvalues of these check operators are $\pm 1$.
 We may therefore choose the stored quantum state $|\Psi \rangle$ as
 a simultaneous eigenstate of 
 all these check operators with the eigenvalue $1$.  

 Let us now consider the case that a chain of phase errors emerge. 
 Bit-flip errors can be treated similarly on the dual lattice.
 We can detect the error chain by operating check operators
 to the stored quantum state.
 Phase error caused by $\sigma^z_l$ can be detected by the check operator
 $X_i$ on the original lattice
 and a chain of bit-flip errors (due to $\sigma^x_l$) 
 by the operator $Z_P$
 on the dual lattice.
 Since we prepare the original state whose eigenvalues of check operators
 are all positive ($1$),
 we can find the end positions of the error chain
 from the positions of a wrong sign ($-1$) of check operators, called syndrome.
 It should be remembered here that we can detect only both ends 
 of the error chain, and we are asked to infer the actual configuration of
 the chain from
 its ends. In some cases our inference of error chain is successful,
 while in other cases we might infer a wrong error chain.
 An example of error correction procedure is shown in Fig. 2 (right).
 We can  detect only the ends (black circles, syndrome) of a real error chain
 ($E$) as the sites where $X_i|\Psi \rangle=-|\Psi \rangle$ in the case of phase errors.
 An example of correction chain inferred from the syndrome
  is shown dotted ($E'$). 
 The example shown in Fig. 2 is a case of successful correction, where
 two chains $E$ and $E'$ are different but homologically equivalent.
 The reason will be explained later.
 
\begin{figure}
\begin{center}
  \includegraphics[width=11cm]{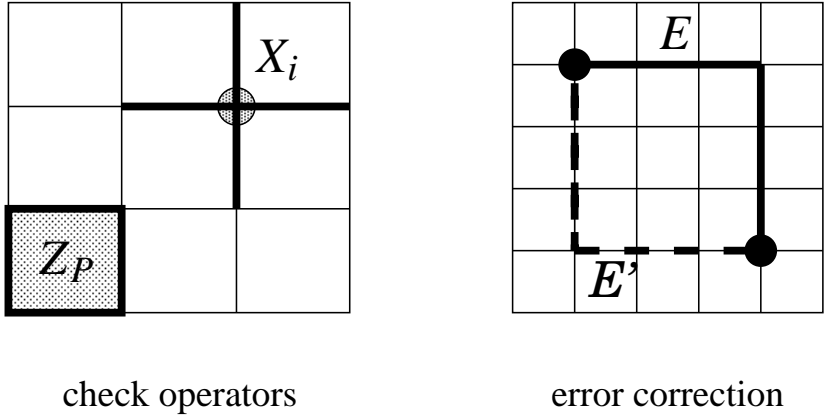}
\label{toriccode2}
\caption{Check operators and the inference procedure of error
 chain. Black circles denote syndrome.}
\end{center}
\end{figure}

In Fig. 3 are shown two examples of error correction.
 An example of successful correction is shown on the left. In this 
 case, the real error chain ($E$, solid line)
 and correction chain ($E'$, dotted line) form a loop, which
 is homologically trivial (or can be shrunk to a
 point). Since the change of $E$ to $E'$ (or vice versa) is represented
 by the operators $Z_p$ on the plaquettes inside the loop $E+E'$ and
 $|\Psi \rangle$ is invariant by their operation, the difference between
 $E$ and $E'$ is not essential,
 \begin{equation}
    \bigotimes_{l \in E}  \sigma^z_l |\Psi \rangle =
    \bigotimes_{l \in E'}  \sigma^z_l |\Psi \rangle.
 \end{equation}
 An example on the right of Fig. 3 is an unsuccessful 
 case. In this case, $E$ and $E'$ form a homologically nontrivial loop on the torus.
 It is impossible to change $E$ to $E'$ by operating $Z_p$'s. 
 Thus an error correction by using $E'$ (dotted lines) yields a different
 (erroneous) result.
 Such an unsuccessful correction procedure
 often occurs when the error rate becomes large
 because many/long error chains may exist.
 It is expected from these examples
 that the accuracy threshold
 of error correction exists.

\begin{figure}
\begin{center}
  \includegraphics[width=13cm]{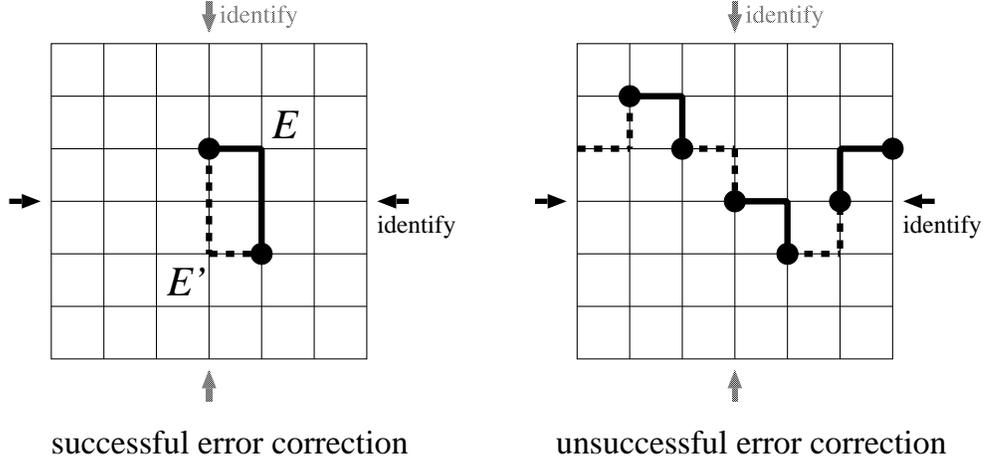}
\label{toriccode3}
\caption{Successful and unsuccessful error corrections.}
\end{center}
\end{figure}
 Next we explain the relation between the $2D$ toric code and the $2D$ RBIM
 (Fig. 4) \cite{DKLP}.
 First let us consider the procedure of successful error correction
 when the real error chain $E$ and
 correction chain $E'$ constitute a homologically trivial loop.
 It is useful to consider the dual lattice.
 We assign a spin on each dual site and an interaction
 for each dual bond (Fig. 4, right).  In doing so,
 we reverse spins inside the loop $E+E'$ and 
 assign antiferromagnetic (reversed sign)
 interactions to the bonds which correspond to the error chain $E$.
 By this procedure
 the $2D$ toric code can be seen as a ($\pm J$) RBIM.
 The error rate $1-p$ of the toric code corresponds to the probability of
 antiferromagnetic interaction for each bond.
 The ratio of error probability to non-error probability $(1-p)/p$ is
 identical with the edge Boltzmann
 factor of local interaction corresponding to the ratio of unfavourable and favourable
 spin alignment, $e^{-K}/e^K$ as seen in Fig. 4.
 Thus the $2D$ RBIM under consideration lies on the Nishimori line
 defined by $e^{-2K}=(1-p)/p$ \cite{NL}.
 
 The relation between the phase diagram of the $2D$ RBIM 
 and accuracy threshold of the toric code is explained intuitively as follows \cite{DKLP,WHP}.
 In the magnetically ordered phase of the RBIM, error correction is successful
 since the islands of reversed spins as depicted on the right of Fig. 4 are not extensively
 large in the ordered phase, which implies that the difference between $E$ and $E'$
 is not significant.
 On the other hand, in the disordered phase,
 the error correction fails and encoded information is lost.
 Thus the critical probability of antiferromagnetic bond at the phase boundary
 along the Nishimori line (multicritical point)
 gives the accuracy threshold of error correction
 in the toric code.
\begin{figure}
\begin{center}
  \includegraphics[width=12cm]{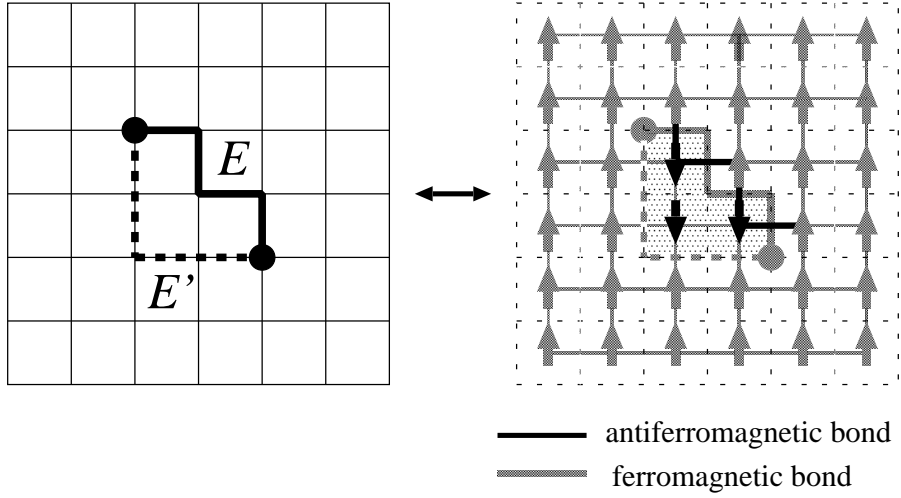}
\label{toriccode4}
\caption{Correspondence between the toric code and the RBIM.}
\end{center}
\end{figure}

 Now we generalize the toric code from the $2D$ bond-qubit system to the $4D$
 plaquette-qubit system.  A motivation is that
 the accuracy threshold of the $4D$ plaquette-qubit system can be estimated
 from the study of the corresponding $4D$ gauge model
 which can be analyzed by duality, as shown in the
 following sections.
 For this purpose we prepare a $4D$ hypercubic lattice and assign a
 qubit on each
 plaquette (Fig. 5). Bit-flip and phase errors occur on each plaquette.
 We define the check operator as follows,
 \begin{eqnarray}
 X_l & = & \bigotimes_{l \in \partial P} \ \sigma^x_P, \\
 Z_C & = & \bigotimes_{P \in \partial C} \ \sigma^z_P, 
 \end{eqnarray}
 where $X_l$ is defined on each bond and $Z_C$ on each $3D$ cube on
 the $4D$ lattice as shown in Fig. 5 (right).
 We can check a phase error $\sigma^z_P$ by the 
 operator $X_l$ and a bit-flip error $\sigma^x_P$ by $Z_C$.
 In the same way as in the $2D$ case, the phase error $\sigma^z_P$ 
 is detected on the original
 lattice and the bit-flip error $\sigma^x_P$
 on the dual lattice.
 Note that the plaquette model on the $4D$ hypercubic
 lattice is self-dual in the sense that the dual system also carries its degrees of
 freedom on plaquettes as elucidated below.
\begin{figure}
\begin{center}
  \includegraphics[width=16cm]{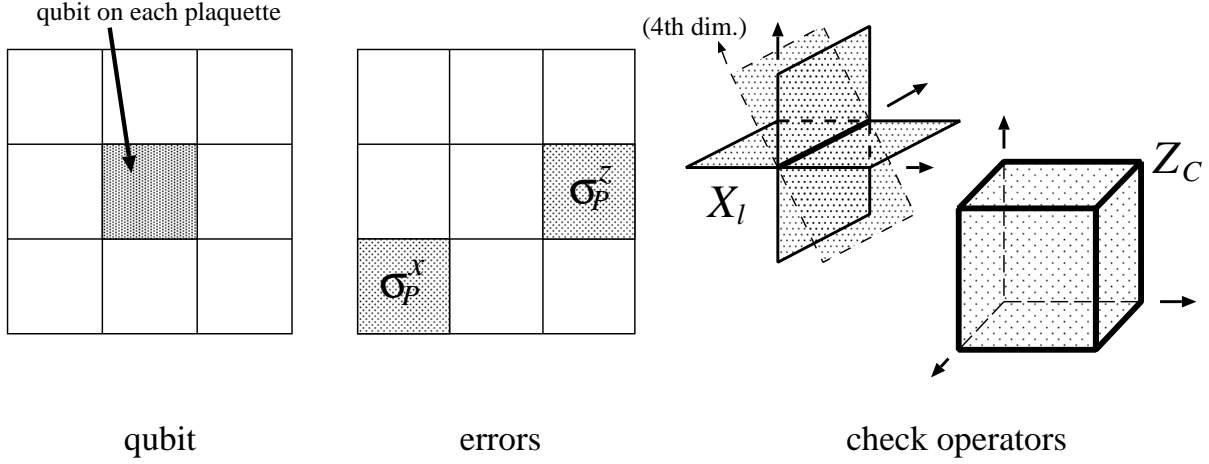}
\label{4Dtoriccode1}
\caption{$4D$ toric code (I):  Qubits errors and check operators
 are displayed. The $2D$ square lattice is a part of the $4D$ hypercubic lattice.}
\end{center}
\end{figure}

 Let us consider
 the correction procedure of phase errors (Fig. 6).
 An error chain in the $2D$ bond toric code becomes an ``error sheet'' in
 the present $4D$ case. 
 We detect the error sheet using the check operator
 $X_l$. We can determine only the boundary of the error sheet from syndrome, which
 is similar to the $2D$ bond case. We must infer the
 error sheet itself from its boundary. 
 In the correction procedure,
 the real error sheet $E$ and the correction sheet $E'$ form a connected
 surface.
 If the surface is closed or homologically trivial
 (middle in Fig. 6),
 error correction is successful because the operations of correcting 
 operators on $E$ and $E'$ are equivalent to each other.
 On the other hand the correction procedure 
 is not successful when the surface is
 homologically nontrivial (right in Fig. 6).
\begin{figure}
\begin{center}
  \includegraphics[width=16cm]{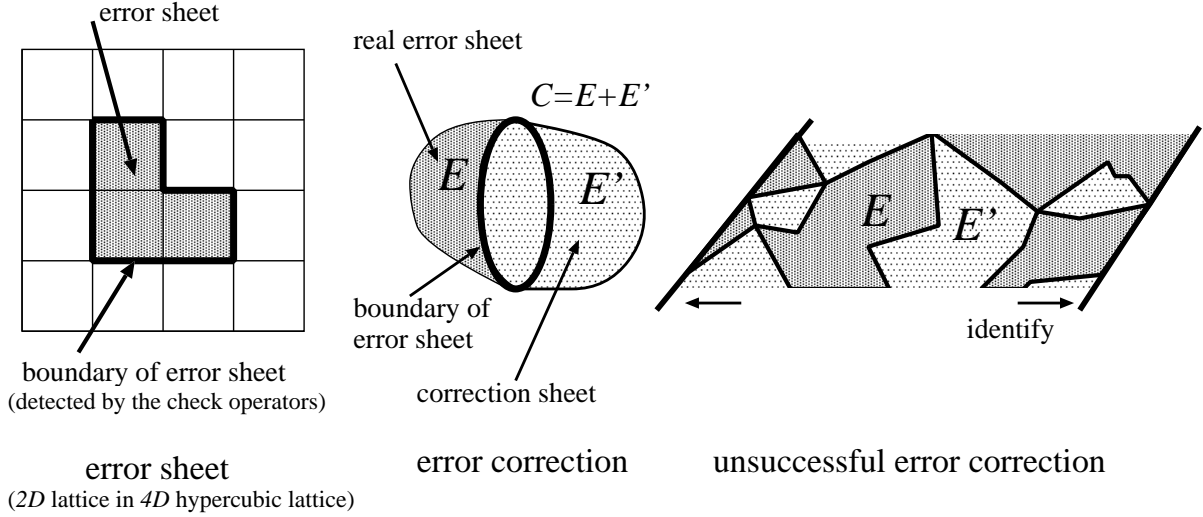}
\label{4Dtoriccode2}
\caption{The $4D$ toric code (II): Error sheet and error correction procedure.
 The error sheet and correction sheet
 shown in the figure are hyperplanar surfaces.}
\end{center}
\end{figure}

 The correspondence of the $4D$ plaquette toric code and the $4D$ RPGM is similar
 to the $2D$ case \cite{DKLP}. 
 Here we consider the case of successful error correction.
 Suppose that an error occurs with probability $1-p$ on each plaquette.
 The probability $\mbox{Prob}(E,E')$ that the error sheet $E$
 and the correction sheet $E'$ are generated is given by (similarly to Fig. 4), \begin{equation}
 \label{probability}
 \mbox{Prob}(E,E') \propto \prod_P \exp(K_P U_P),
 \end{equation}
 where $U_P$ takes $-1$ on the surface $C=E+E'$ and $1$ elsewhere. Here 
 $K_P$ is defined by
 \begin{equation}
 e^{-2 K_P} = \left\{ \begin{array}{c} 
 (1-p)/p, \ \ \mbox{for} \ \ P \hspace{0.3mm} \in \hspace{-2.6mm} | \hspace{1.6mm}  E, \\
 p/(1-p), \ \ \mbox{for} \ \ P \in E.
 \end{array}
 \right.
 \end{equation}
 For each bond $l$ on the $4D$ lattice, $U_P$ must satisfy the constraint
 (because a connected surface $C$ has no boundary),
 \begin{equation} 
 \prod_{P: \ l \in \partial P} U_P =1,
 \end{equation}
 or on the dual lattice,
 \begin{equation} 
 \prod_{P': \ P' \in \partial C'} U_{P'} =1,
 \end{equation}
 for each dual cube $C'$.
 We introduce dual bond variables to solve this constraint,
 \begin{equation}
 \label{plaqueeteval}
 U_{P'}\ ( =U_{P} )=  \prod_{l' \in \partial P'} \tilde{\xi}_{l'},
 \end{equation}
 where $\tilde{\xi}_{l'}$ is an ordinary Ising variable which takes
 $\pm 1$.
 Using $K_{P'} (= K_P)$ and $U_{P'}$,
 we can rewrite the probability of
 Eq. (\ref{probability}),
 \begin{equation}
 \mbox{Prob}(E,E') \propto \prod_{P'} \exp\left\{ K_{P'} U_{P'}(\tilde{\xi}_{l'}) \right\}
 =  \exp\left\{ K \sum_{P'} \tau_{P'} U_{P'}(\tilde{\xi}_{l'}) \right\},
 \end{equation} 
 where $\tau_{P'} = K_{P'} / K $ takes $\pm 1$.
 If we take the sum over $\tilde{\xi}_{l'}$,
 This turns to the partition
 function of the $4D$ RPGM, which we will study in Sec. 5.
 $K$ is defined by 
 $e^{-2K}=(1-p)/p$, the condition of the Nishimori line \cite{NL}.
 
 \section{2$D$ Ising model and Wu-Wang duality}
 
 The next step is to develop a framework to investigate the mathematical
 structure of the models defined in Sec. 2.
 We consider the $2D$ Ising model on the $2D$ square lattice
 in order to illustrate the powerful technique of Wu-Wang duality.
 The Hamiltonian of the Ising model is
 \begin{equation}
 \label{non-randommodelising}
 H= -J \sum_{\langle ij \rangle} \tilde{S}_i \tilde{S}_j,
 \end{equation}
 where $\tilde{S}_i$ is the normal Ising spin variable which takes $\pm 1$ and
 $J$ is the uniform coupling constant. This Hamiltonian can be rewritten as,
 \begin{equation}
 \label{non-randommodelising2}
 H = J \sum_{l} \xi_{l}, \hspace{5mm} \mbox{} \ \ \xi_{l} =  \sum_{i \in \partial l} S_{i},
 \end{equation}
 where $l$ and $i$ are bond and site,
 respectively, and $S_{i}$ is the modulo-2 spin
 variable which takes 0 or 1 in this case. The summation over $i$ in the
 definition of $\xi_l$ is also defined by modulo 2. Note that a product
 of $\tilde{S}_i$'s has been changed to a sum of $S_i$'s. 
\begin{figure}
\begin{center}
  \includegraphics[width=12cm]{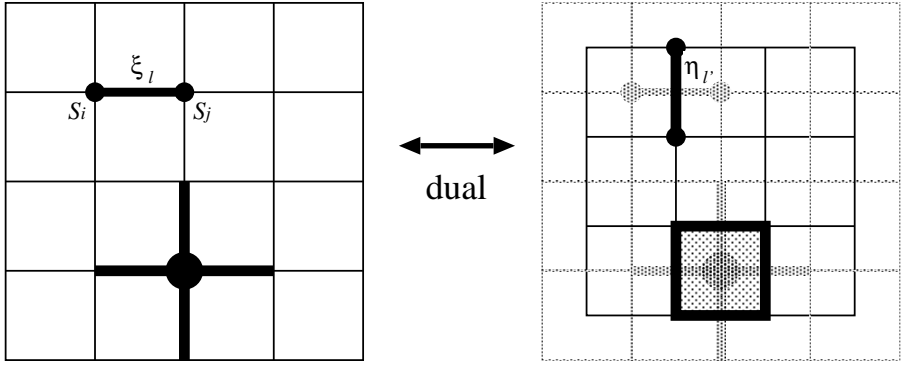}
\label{figisingdual}
\caption{The $2D$ Ising model and its self-duality. Also illustrated is the
 derivation of modulo-2 Kronecker delta of  Eq. (\ref{dualsum0ising}) for
 each plaquette.}
\end{center}
\end{figure}

 The partition function is,
 \begin{equation}
 Z = \sum_{S_{i}= 0,1} \prod_{l} u(\xi_l), \hspace{5mm} \mbox{} \ \
 u(\xi_l) = \exp [ -K \xi_l (S_{i}) ],
  \label{bondsumising}
 \end{equation}
 where $K= \beta J=J/kT$.
 To take the dual of this model, 
 we introduce a function obtained by regarding the modulo-2 bond variables
 $\xi_{l}$ as independent variables in Eq. (\ref{bondsumising}),
 \begin{equation}
 \label{plasumising}
 \tilde{Z} = {\sum_{\xi_{l} = 0,1}} \prod_{l} u(\xi_l), \hspace{5mm} \mbox{} \ \ u(\xi_l) = \exp [ -K \xi_{l} ].
 \end{equation}
 Note that two functions Eqs. (\ref{bondsumising}) and
 (\ref{plasumising}) do not coincide in their naive forms.
 To establish their equivalence,
 we must impose the following condition on the
  summation Eq. (\ref{plasumising}),
 \begin{equation}
 \label{cond1ising}
 \sum_{l \in \partial P} \xi_{l} = 0 \hspace{3mm} \mbox{(mod 2)}  \hspace{5mm} \mbox{ for all $P$},
 \end{equation}
 where $P$ is a plaquette on the $2D$ square lattice,
 because $\xi_l$ is composed of the modulo-2 sum of two $S_i$'s at
 the ends of $l$. 
 This means that change of the value of spin $S_{i}$ at an arbitrary site
 on a $2D$ plaquette always causes modification two of
 the four $\xi_{l}$'s in the plaquette,
 and l.h.s. of Eq. (\ref{cond1ising}) for any plaquette
 remains zero modulo 2.
 We denote the summation with this condition by a prime.
 For Eq. (\ref{plasumising}), 
 \begin{equation}
 \label{plasum2ising}
 Z(u) = \mbox{const.} \times {\sum_{\xi_{l} = 0,1}}^{\!\!\! '} \prod_{l} u(\xi_l),
 \end{equation}
 and we may then
 identify the two $Z$'s in Eqs. (\ref{bondsumising}) and (\ref{plasum2ising}).

 The dual representation of the partition function is derived as follows.
 The dual of a bond $l$ ($1D$ object)
 on the $2D$ square lattice is also a bond $l'$ (Fig. 7).
 We define the dual bond variables $\eta_{l'}$ 
 on the dual lattice, which also take the value $0$ or $1$,
 and perform a discrete Fourier transformation of $u(\xi_l)$ defined by
 \begin{equation}
 \label{fourier0ising}
 u^*(\eta_{l'})  =  
 \frac{1}{\sqrt{2}} \sum_{\xi_{l}=0,1} \exp(\pi i \xi_{l} \eta_{l'}) u(\xi_{l}) 
  =  \frac{1}{\sqrt{2}}\sum_{\xi_{l}=0,1} 
  \exp \left( \pi i \left(\sum_{i \in \partial l} S_{i} \right) \eta_{l'} \right) u(\xi_{l}),
 \end{equation}
 or conversely,
 \begin{equation}
 \label{fourierising}
 u (\xi_{l}) = 
 \frac{1}{\sqrt{2}} \sum_{\eta_{l'}=0,1}
  \exp \left( \pi i \left(\sum_{i \in \partial l} S_{i} \right) \eta_{l'} \right) u^*(\eta_{l'}). 
 \end{equation}
 More explicitly, the edge Boltzmann factors are,
 \begin{eqnarray}
 u(0) &=& 1, \nonumber \\
 u(1) &=& e^{-K}, \nonumber \\
 u^* (0) &=& \frac{1}{\sqrt{2}}(1+e^{-K}), \nonumber \\
 u^* (1) &=& \frac{1}{\sqrt{2}}(1-e^{-K}).
 \end{eqnarray}
 We use Eq. (\ref{fourierising}) in Eq. (\ref{bondsumising}) and take the sums
 over $S_i$.  
 Considering that a single site always appears at the 
 ends of four bonds on the $2D$ square lattice, 
 the exponential factors in Fourier transformation
 lead (modulo-2) to Kronecker deltas such as,
 \begin{equation}
 \label{dualsum0ising}
 \delta_{\rm mod~2} \left(\sum_{l' \in \partial P'} \eta_{l'} \right)  \hspace{1cm}
 \mbox{ for all $P'$},
 \end{equation}
 where $P'$ is the dual plaquette on the $2D$ dual square lattice (Fig.7).
 This condition is equivalent to Eq. (\ref{cond1ising}) and
 we can replace the summation over $\eta_{l'}$ with the constrained one.
 
 Now we can express the partition function using dual variables,
 \begin{equation}
 \label{dualplasumising}
 Z(u^*) = \mbox{const.} \times {\sum_{\eta_{l'} = 0,1}}^{\!\!\! '} \prod_{l'} u^*(\eta_{l'}).
 \end{equation}
 We have therefore obtained a direct relation between the two partition functions in
 Eqs. (\ref{plasum2ising}) and (\ref{dualplasumising}),
 \begin{equation}
 \label{dualZising}
 Z(u) = \mbox{const.} \times Z(u^*).
 \end{equation}
 The transition point, if unique, is identified with the fixed point
 of duality transformation of the edge Boltzmann factor,
 \begin{equation}
 u(0) = u^{*}(0), \hspace{5mm} u(1) = u^{*}(1),
 \end{equation}
 both of which lead to the same relation\footnote{If we start with the
 Hamiltonian Eq.(\ref{non-randommodelising}), the transition
 point will be given by the relation $e^{-2K_c} = \sqrt{2} -1$. This 
 difference is caused by the definitions of the
 coupling $J$ in Eqs. (\ref{non-randommodelising}) and
 (\ref{non-randommodelising2}), which differ by factor 2.}
 \begin{equation}
 \label{criticalpoint}
 e^{-K_c} = \sqrt{2} -1.
 \end{equation}
 
 \section{4$D$ $Z_{2}$ lattice gauge model and Wu-Wang duality}
 
 Let us move on to the $4D$ $Z_{2}$ gauge model.
 We can treat this model using duality in an analogous manner
 to the $2D$ Ising model.
\begin{figure}
\begin{center}
  \includegraphics[width=5cm]{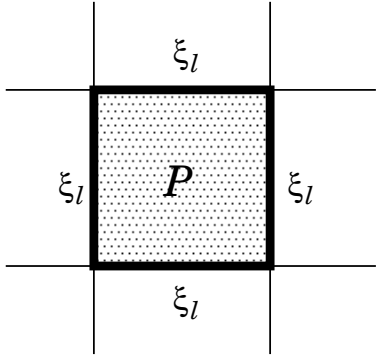}
\label{figexample1}
\caption{Bond variables and a plaquette.}
\end{center}
\end{figure}
 The Hamiltonian is
 \begin{equation}
 \label{non-randommodel}
 H = J \sum_{P} U_{P}, \hspace{5mm} \mbox{} \ \ U_P = 
  \sum_{l \in \partial P} \xi_{l} \hspace{5mm} 
  (\simeq -\prod_{l \in \partial P} \tilde{\xi}_{l}),
 \end{equation}
 where $P$ and $l$ are the plaquette and bond on
 the $4D$ hypercubic lattice,
 respectively (Fig. 8),
 and $J$ is the uniform coupling constant.
 Here $\xi_{l}$ is the modulo-2 variable which takes 0 or 1 and the sum
 over $l$ is also defined by modulo 2. 
 ($\tilde{\xi}_{l}$ is the ordinary Ising spin variable which takes
 $\pm 1$.)
 The partition function is,
 \begin{equation}
 Z = \sum_{\xi_{l}= 0,1} \prod_{P} u(U_P), \hspace{5mm} \mbox{} \ \
 u(U_P) = \exp [ -K U_P (\xi_{l}) ].
  \label{bondsum}
 \end{equation}
 To take the dual, 
 we regard $U_P$ as the independent plaquette variables,
 \begin{equation}
 \label{plasum}
 \tilde{Z} = {\sum_{U_P = 0,1}} \prod_{P} u(U_P), \hspace{5mm} 
 \mbox{} \ \ u(U_P) = \exp [ -K U_P ].
 \end{equation}
 Functions (\ref{bondsum}) and
 (\ref{plasum}) do not necessarily coincide unless
 we impose the following condition on the summation Eq. (\ref{plasum}),
 \begin{equation}
 \label{cond1}
 \sum_{P \in \partial C} U_P = 0 \hspace{3mm} \mbox{(mod 2)}  \hspace{5mm} \mbox{ for all $C$},
 \end{equation}
 where $C$ is a $3D$ cube on the $4D$ hypercubic lattice. 
 This means that, if we change the value of $\xi_{l}$ at an arbitrary edge
 of the $3D$ cube, two of six $U_P$'s on the faces automatically change,
 and l.h.s. of Eq. (\ref{cond1}) for any cube
 remains zero modulo 2 (Fig. 9).
\begin{figure}
\begin{center}
  \includegraphics[width=10cm]{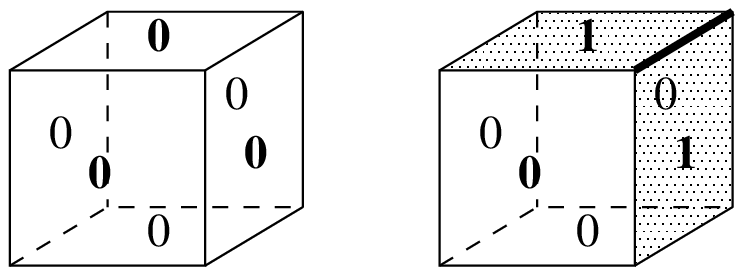}
\label{figexample2}
\caption{Illustration of the condition Eq. (\ref{cond1}): A $3D$ cube and
 $U_P$'s of six faces are shown. The operation  of
 changing $\xi_{l}$\ for one edge (shown in a thick line) always induces
 the changes of two $U_P$'s.}
\end{center}
\end{figure}
 We denote the summation with this condition with a prime.
 For Eq. (\ref{plasum}), 
 \begin{equation}
 \label{plasum2}
 Z(u) = \mbox{const.} \times {\sum_{U_{P} = 0,1}}^{\!\!\! '} \prod_{P} u(U_P).
 \end{equation}
 This expression is identified with Eq. (\ref{bondsum}).

 Next we represent the partition function in the dual form.
 The dual of a plaquette $P$ ($2D$ object)
 on the $4D$ hypercubic lattice is also a plaquette $P'$.
 We define the variables $V_{P'}$ 
 on dual plaquettes, which also take the value $0$ or $1$,
 and perform a Fourier transformation of $u(U_P)$,
 \begin{equation}
 \label{fourier0}
 u^*(V_{P'})  =  \frac{1}{\sqrt{2}} \sum_{U_{P}=0,1} \exp(\pi i U_{P} V_{P'}) u(U_{P}) 
  =  \frac{1}{\sqrt{2}}\sum_{U_{P}=0,1}
   \exp\left(\pi i \left(\sum_{l \in \partial P} \xi_{l} \right) V_{P'}\right) u(U_{P}),
 \end{equation}
 or conversely,
 \begin{equation}
 \label{fourier}
 u (U_{P}) = \frac{1}{\sqrt{2}} \sum_{V_{P'}=0,1} 
 \exp\left(\pi i \left(\sum_{l \in \partial P} \xi_{l} \right) V_{P'}\right) u^*(V_{P'}). 
 \end{equation}
 We insert Eq. (\ref{fourier}) into Eq. (\ref{bondsum}) and take the sums
 over $\xi_l$.  
 Considering that a bond is always at the
 boundaries of six faces, 
 the exponential factors in the Fourier transformation
 yield deltas (modulo-2) such as,
 \begin{equation}
 \label{dualsum0}
 \delta_{\rm mod~2} \left(\sum_{P' \in \partial C'} V_{P'} \right)  \hspace{1cm}
 \mbox{ for all $C'$},
 \end{equation}
 where $C'$ is the dual cube on the $4D$ dual hypercubic lattice (Fig. 10).
\begin{figure}
\begin{center}
  \includegraphics[width=11cm]{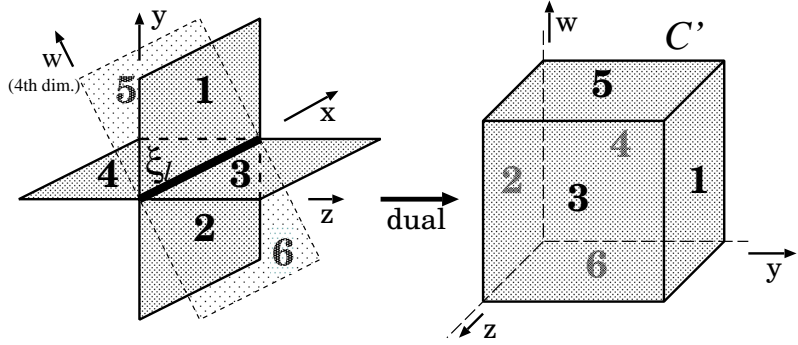}
\label{figexample3}
\caption{Illustration of the condition Eq. (\ref{dualsum0})
 in the dual space:
 $3D$ cube in the dual space is localized in the $x$ direction.}
\end{center}
\end{figure}
 This condition is equivalent to Eq. (\ref{cond1}) and we can replace the
 summation over $V_{P'}$ with the constrained one.
 
 We have thus obtained the partition function using the dual variables,
 \begin{equation}
 \label{dualplasum}
 Z(u^*) = \mbox{const.} \times {\sum_{V_{P'} = 0,1}}^{\!\!\! '} \prod_{P'} u^*(V_{P'}).
 \end{equation}
 The two partition functions in
 Eqs. (\ref{plasum2}) and (\ref{dualplasum}) are
 related as,
 \begin{equation}
 \label{dualZ}
 Z(u) = \mbox{const.} \times Z(u^*).
 \end{equation}
 From this relation, we can derive the transition point as in the $2D$ case,
 \begin{equation}
 e^{-K_c} = \sqrt{2} -1,
 \end{equation}
 which is equivalent to the one of the 2D Ising model Eq.(\ref{criticalpoint}).

\section{4$D$ random plaquette gauge theory}
 Next we explain the technique of dual transformation
 to deal with the RPGM following  Ref. \cite{MNN}.
 They studied the $2D$ ($\pm J$) RBIM mainly using the Wu-Wang duality.
 Our method for analyzing the $4D$ RPGM is essentially the same as in Ref. \cite{MNN}.
 
 From the motivation related to quantum error correction explained in 
 Sec. 2,
 we incorporate randomness in the 
 plaquette gauge model Eq. (\ref{non-randommodel}),
 \begin{equation}
 H = J \sum_{P} \tau_{P} U_{P},
 \end{equation}
 where $\tau_{P}$ is a plaquette-dependent quenched random variable.
 Here we assume that $\tau_{P}$ takes $1$ with
 probability $p$ and $-1$ with $1-p$.
 
 We make use of the replica technique for averaging over randomness in
 $\tau_{P}$. That is, we prepare
 $n$ replica systems and take the $n \rightarrow 0$ limit in the end
 to obtain physical quantities. 
 After the configuration average over random plaquette variables,
 the partition function is determined by local non-random plaquette Boltzmann
 factors because the system then acquires spatial homogeneity.
 In the following
 we define the ``averaged'' plaquette Boltzmann factor for a given
 probability $p$.
  
 First we consider the simplest case of $n=1$.
 We define the plaquette Boltzmann matrix\footnote{In this
 section and Sec. 6 we define the
 Boltzmann factor by the normal Ising spin variable $\tilde{\xi}_l$
 in Eq. (\ref{non-randommodel}) instead of the modulo-2 variable $\xi_l$
 in order to follow the notation of Ref. \cite{MNN}.} as
 \begin{equation}
 \label{1replica}
  p
 \left[
 \begin{array}{cc}
 \kappa_+ & \kappa_- \\
 \kappa_- & \kappa_+
 \end{array}
 \right]
 + (1-p)
 \left[
 \begin{array}{cc}
 \kappa_- & \kappa_+ \\
 \kappa_+ & \kappa_-
 \end{array}
 \right]
 \equiv
 \left[
 \begin{array}{cc}
 x_0 & x_1 \\
 x_1 & x_0
 \end{array}
 \right]
 = A_1,
 \end{equation}
 where $\kappa_{\pm}=e^{\pm K}$. 
 $x_0$ means the averaged Boltzmann factor for the configuration  
 $\prod_{l} U_{l}$ = 1 and $x_1$ for  $\prod_{l} U_{l} = -1$.
 For a generic $n$,
 we can obtain the $2^{n} \times 2^{n}$ matrix $A_n$ by a recursive procedure,
 \begin{equation}
 \label{onereplica}
  p
 \left[
 \begin{array}{cc}
 \kappa_+ & \kappa_- \\
 \kappa_- & \kappa_+
 \end{array}
 \right]^{\otimes n}
 + (1-p)
 \left[
 \begin{array}{cc}
 \kappa_- & \kappa_+ \\
 \kappa_+ & \kappa_-
 \end{array}
 \right]^{\otimes n}
 \equiv
 A_n
 = \left[
 \begin{array}{cc}
 A_{n-1} & B_{n-1} \\
 B_{n-1} & A_{n-1}
 \end{array}
 \right],
 \end{equation}
 where $B_{n}$ is obtained from $A_{n}$ by replacing $x_k \rightarrow
 x_{k+1}$. The factor $x_k$ is defined by generalizing $x_0$ and $x_1$
 in Eq. (\ref{1replica}) and
 corresponds to the averaged plaquette Boltzmann factor 
 for the configuration $\prod_l U_l =1$ in $n-k$ replicas and $-1$ in $k$ replicas.
 The explicit form of $x_k$ is
 \begin{equation}
 x_k = p {\kappa_+}^{n-k}{\kappa_-}^{k} + (1-p) {\kappa_+}^{k}{\kappa_-}^{n-k}.
 \end{equation} 
 The averaged partition function depends on averaged plaquette Boltzmann
 factors $x_k$,
 \begin{equation}
 \label{partdual}
 [Z^n]_{\rm{av}} \equiv  Z_n (x_0,x_1,\ldots, x_n),
 \end{equation}
 where $[\ ]_{\rm{av}}$ means random average. 

 On the dual lattice, we can also define the dual averaged plaquette Boltzmann
 factor ${x_k}^{*}$ similarly.
 The Boltzmann factors for 
 the original and dual systems are related by Eq. (\ref{fourier0}), and accordingly
 the relations between $x_k$ and $x_k^*$ for $n=1$ are given by
 replacing $u$ and $u^*$ with $x$ and $x^*$ respectively, 
 \begin{eqnarray} 
 \label{dualfactor}
 \sqrt{2} x_0^{*} = x_0 + x_1, \nonumber \\
 \sqrt{2} x_1^{*} = x_0 - x_1.
 \end{eqnarray} 
 For $n=2$, 
 \begin{eqnarray}
 2 x_{0}^{*} &=& (x_0 + x_1) + (x_1 + x_2) = x_0 + 2 x_1 + x_2, \nonumber \\
 2 x_{1}^{*} &=& (x_0 - x_1) + (x_1 - x_2) = x_0 - x_2, \nonumber \\
 2 x_{2}^{*} &=& (x_0 - x_1) - (x_1 - x_2) = x_0 - 2 x_1 + x_2, 
 \end{eqnarray}
 similarly to the $2D$ case \cite{MNN}.

 In a similar way, 
 we can obtain the explicit form of $x_{k}^{*}$ for $n$ replicas,
 \begin{eqnarray}
 2^{n/2} x_{2m}^* & = &  (\kappa_+ + \kappa_-)^{n-2m} (\kappa_+ - \kappa_-)^{2m}, \nonumber \\
 2^{n/2} x_{2m+1}^* & = & (2p-1) (\kappa_+ + \kappa_-)^{n-2m-1} (\kappa_+ - \kappa_-)^{2m+1}.  
 \end{eqnarray}
 We can write the duality of the partition function using $x_{k}$ and
 $x_{k}^*$,
 \begin{equation}
 \label{selfdual}
 Z_n (x_0,x_1, \ldots x_n) = Z_n (x_0^*,x_1^*, \ldots x_n^*),
 \end{equation}
 where we have neglected the trivial factor in Eq. (\ref{dualZ})
 which is irrelevant to thermodynamic properties.

 This procedure is completely the same as in the $2D$ RBIM.
 In the $2D$ RBIM, the multicritical point on the Nishimori line
 defined by
 \begin{equation}
 \label{NL}
 e^{-2K} = \frac{1-p}{p},
 \end{equation}
 is supposed to give the lower bound of the probability $p$ for
 ferromagnetic (ordered) phase.
\begin{figure}
\begin{center}
\unitlength=1cm
  \includegraphics[width=8cm]{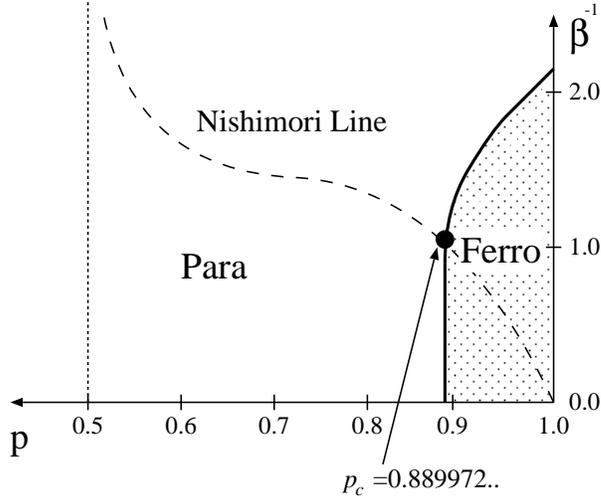}
\label{phasediagram}
\caption{Phase diagram of the $2D$ $\pm J$ RBIM}
\end{center}
\end{figure}
 In Ref. \cite{MNN}, they conjectured that the phase boundary 
 on the Nishimori line coincides with the crossing point of
 Eq. (\ref{NL}) and a line defined by the relation for the 
 averaged Boltzmann factor,
 \begin{equation}
 \label{conjecture}
 x_0 = x_0^*.
 \end{equation} 
 They confirmed that this relation together with the condition of Eq. (\ref{NL}) leads
 to the exact multicritical point at least in the case of
 small and infinite $n$.
   
 We write Eq. (\ref{conjecture}) explicitly with inverse temperature $K$
 and probability $p$,
 \begin{equation}
 p e^{n K} + (1-p) e^{-n K} = 2^{-n/2} (e^K + e^{-K})^n.
 \end{equation}
 In conjunction with the Nishimori relation Eq. (\ref{NL}), this yields,
 \begin{equation}
 \label{probcritical}
 p_c ^{n+1} + (1-p_c)^{n+1} = 2^{-n/2},
 \end{equation}
 where $p_c$ is the probability at the critical point on the Nishimori line
 (Fig. 11).
 If we expand Eq. (\ref{probcritical}) with respect to
 $n$ and take the $n \rightarrow 0$ limit, we obtain the relation of
 order $O(n^1)$ terms,
 \begin{equation}
 \label{probcritical2}
 -p_c \log p_c - (1-p_c) \log (1-p_c) = \frac{\log2}{2}.
 \end{equation}
 This equation gives $p_c=0.889972\ldots$ numerically, which is in good 
 agreement with the numerical studies in Ref. \cite{numerical}.

 This argument for the $2D$ RBIM is also applicable to the $4D$ RPGM. In
 this latter model the order parameter is defined by the Wilson loop operator,
 \begin{equation}
 W[C] = \prod_{l \in C} \tilde{\xi}_{l},
 \end{equation}
 where $C$ is an arbitrary closed loop. The
 random and thermal averaged value of $W[C]$ obeys,
 \begin{equation}
 \ln ( [ W[C] ]_{{\rm av},K} ) \propto \left\{ 
 \begin{array}{cc}
 \hspace{-2.5cm}  [\mbox{length of C}] \hspace{0.5cm} (\mbox{perimeter law})   \\
  \hspace{2.3cm} \mbox{for ordered (Higgs) phase,} \\
 \hspace{-3cm}  [\mbox{area of C}]  \hspace{1cm} (\mbox{area law})  \\ 
  \hspace{2.3cm} \mbox{for disordered (confinement) phase,}
 \end{array} \right. 
 \end{equation}
 where $[\ ]_{{\rm av},K}$ means
 random and thermal averaged quantity.
 Higgs (ordered) and confinement (disordered) phases correspond to 
 ferromagnetic and paramagnetic phase, respectively.
 If we apply the previous discussion for the $2D$ RBIM to
 the $4D$ RPGM, we automatically obtain the phase diagram for the $4D$ RPGM
 from the one for the $2D$ RBIM by replacing the names of phases.
 The transition probability at the critical point on the Nishimori line
 is also given by Eq. (\ref{probcritical2}).
 We therefore conjecture that $p_c=0.889972\ldots$ satisfying Eq. (\ref{probcritical2})
  is the exact accuracy threshold of the $4D$ plaquette toric code
 corresponding to the $4D$ RPGM.

 \section{3$D$ RPGM and dual representation}
 In this section we analyze the dual 
 Boltzmann factor for the $3D$ RPGM 
 and its representation by a spin model with ferromagnetic interactions.
 The discussion follows Ref. \cite{MNN} where they 
 mainly treated the $2D$ replicated $\pm J$ Ising model.

 We consider the $\pm J$ RPGM with the Hamiltonian
 \begin{equation}
 H = -J \sum_{P} \tau_{P} U_{P},
 \end{equation}
 on the $n$-replicated $3D$ cubic lattice.
 The probability distribution of $\tau_p$ is the
 same as in the previous section.
 The averaged Boltzmann factors $x_k$ and the dual
 averaged Boltzmann factors $x_k^*$ are defined similarly
 to Sec. 5. It should be remembered, however, that
 the interpretation of dual Boltzmann factor is different in this case;
 In the $3D$ system the plaquette model is not self-dual (that is,
 the duality relation of the partition 
 function in Eq. (\ref{selfdual}) does not hold), and 
 the dual of a plaquette ($2D$ object) is a bond ($1D$ object)
 as illustrated in Fig. 12.
 If we consider the averaged Boltzmann factor $x_k$ on a plaquette on the
 original lattice,
 its dual Boltzmann factor $x_k^*$ for a dual bond $l^*$ is defined
 for the configuration $S^*_i S^*_j =1 \ (S^{*}_{i} $ is a dual spin
 variable and $i,j \in \partial l^*)$ in $n-k$
 replicas and $S^*_i S^*_j =-1$ in the remaining $k$ replicas.
 
 It is instructive to take the ratio of dual Boltzmann factors to $x_0^*$,
 \begin{eqnarray}
 x^{*}_{2m-1} / x^{*}_{0} & = & (2p-1) \left( \frac{\kappa_{+} - \kappa_{-}}{\kappa_{+} + \kappa_{-} } \right)^{2m-1} = (2p-1) \tanh^{2m-1} K, \nonumber \\
 x^{*}_{2m} / x^{*}_{0} & = & \tanh^{2m} K. 
 \end{eqnarray} 
 This ratio can be written in the simple form
 \begin{equation}
 \label{3Ddualfactor}
 \exp \{ K^* (S^{*(1)}+S^{*(2)}+ \ldots + S^{*(n)}) + K_p^* S^{*(1)} S^{*(2)}\ldots S^{*(n)} \} \equiv Z_n(K,K_p) ,
 \end{equation}
 where $\tanh K = e^{-2K^*}$ and $2p-1 = e^{-2K_p^*} ( \equiv \tanh K_p)$.
 $S^{*(k)}$ is the Ising
 interaction factor in the $k$th replica,
 \begin{equation} 
  S^{*(k)} = S_i^{*(k)} S_j^{*(k)} \hspace{7mm} (i,j \in \partial l^*).
 \end{equation} 
 From the Boltzmann factor Eq.(\ref{3Ddualfactor}), the dual model can
 be interpreted as a spin model
 which has ferromagnetic Ising interactions in each replica
 and an interaction between replicas. 
 
 Next we take the $K_p^* \rightarrow \infty \ (K_p \rightarrow 0)$ limit which corresponds to
 the $p=1/2$ case. In this case $x^*_{2m-1}/x^*_0$ are zero for any $m$ and
 $S^{*(1)}S^{*(2)}S^{*(3)} \ldots S^{*(n)}$ is unity. Therefore
 a spin variable can be removed
 by the relation $S^{*(n)} = S^{*(1)}S^{*(2)} \ldots S^{*(n-1)}$ 
 and the dual Boltzmann factor becomes
 \begin{eqnarray}
 \label{3Ddualfactor2}
 & &  Z_{n} (K, K_p=0) \nonumber \\
 & = &  A \exp \{ K^* (S^{*(1)}+S^{*(2)}+ \ldots + S^{*(n-1)} +  S^{*(1)} S^{*(2)} \ldots S^{*(n-1)} ) \} \nonumber \\
 & = & Z_{n-1}(K,K_p=K). 
 \end{eqnarray}  
 Thus the $n$-replicated system with $K_p=0$ ($p=1/2$) is equivalent to the
 $(n-1)$-replicated system with $K=K_p$ (on the Nishimori line).
 The Boltzmann factor Eq. (\ref{3Ddualfactor}) includes only
 ferromagnetic interactions and the Griffiths inequality
 \cite{Griffiths}
 holds, which leads to monotonicity of order parameters
 with respect to the change of $K$ and $K_p$ ($T$ and $p$). In particular the
 phase boundary between ferromagnetic and non-ferromagnetic phases
 is found to be a monotonic function of $T$ and $p$ (which are
 the parameters of the original model(=$3D$ RPGM)).
 This fact yields,
 \begin{equation}
 T_c^{n} (p=1/2) \ ( =  T_c^{n-1} (\mbox{MCP}) ) \ \leq T_c^{n} (\mbox{MCP}),
 \end{equation}
 where $T_c^{n}$ is the critical temperature for the $n$-replicated system and
 MCP means the multicritical point. We rewrite this inequality using
 probability $p$,
 \begin{equation}
 p_c^{n-1} (\mbox{MCP}) \geq p_c^{n} (\mbox{MCP}),
 \end{equation}
 where $p_c^{n}$ is the probability of the multicritical point for the 
 $n$-replicated system.
 Let us suppose that this inequality holds in the
 $n \rightarrow 0$ limit. 
 If we consider the $n=1$ system on the Nishimori line $(K^* = K^*_p)$,
 the dual Boltzmann factor becomes,
 \begin{equation}
 \label{3Ddualfactor3}
  Z_{1} (K, K_p=K) =  A \exp \{ 2 K^* S^{*(1)} \}. 
 \end{equation} 
 This is the simple ferromagnetic Ising Boltzmann factor and we can 
 make use of the knowledge of the $3D$ ferromagnetic Ising model for the
 analysis.
 The critical point of the $3D$ non-random ferromagnetic Ising model
 is estimated as $2K^* \simeq 0.22165 \ldots$ numerically \cite{3DIsing}.
 Using the condition
 ($K=K^*$) and the dual relation ($2p-1 = e^{-2K_p^*}$), we obtain
 the value $p_c^{1} (\mbox{MCP}) \simeq 0.9006$.
 Finally we obtain the following inequality,
 \begin{equation}
 p_c^{0} (\mbox{MCP}) \geq p_c^{1} (\mbox{MCP}) \simeq 0.9006.
 \end{equation}
 This inequality gives a lower bound of probability at the
 multicritical point for the $3D$ RPGM under the assumption that
 the inequality holds in the limit $n \rightarrow 0$.
  
 The $3D$ RPGM is related to the toric code as well.
 We consider the toric code on the $3D$ cubic lattice and locate
 a qubit on each bond.
 If we choose check operators for this system
 as shown in Fig.12 (bond case), the accuracy threshold of {\it phase} error
 correction is given by
 the probability at phase boundary (on the Nishimori line) in the $3D$ RPGM.
 The accuracy threshold of {\it bit-flip} error correction 
 is given by the phase boundary in the $3D$ RBIM.
 On the other hand, if
 we locate a qubit on each plaquette and choose the check operators
 like in Fig. 12 (plaquette case), the accuracy threshold of {\it bit-flip}
 error correction is given by the probability at phase boundary
 (on the Nishimori line) in the $3D$ RPGM.
 Similarly, the accuracy threshold of {\it phase} error correction is given by
 the critical point in the $3D$ RBIM.
 From this correspondence, we easily find out that the
 procedure of phase error correction in the $3D$ bond qubit system
 and that of bit-flip error correction 
 in the $3D$ plaquette qubit system are related by duality. 
 In the same way, the bit-flip error correction in the $3D$ bond system and 
 phase error correction in the $3D$ plaquette system are dual to each other.
\begin{figure}
\begin{center}
  \includegraphics[width=11cm]{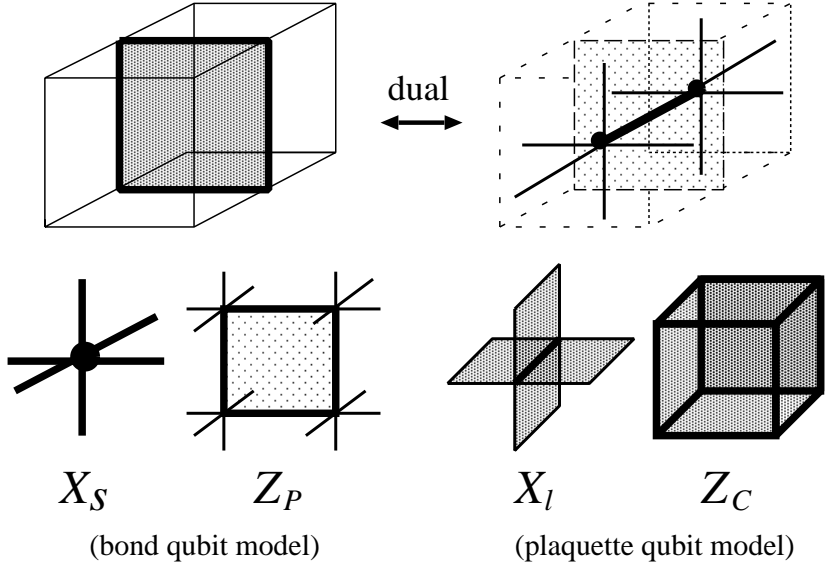}
\label{3Dduality}
\vspace{-0mm}
\caption{Duality of the $3D$ cubic lattice and check operators: $X_s (X_l)$ is
 given by the product of Pauli matrix $\sigma_x$ for each bond
 (plaquette).
 $Z_P (Z_C)$ is given by the product of $\sigma_z$ in the same way.}
\end{center}
\end{figure}

 \section{Conclusion and discussion}  
 We have discussed the phase structure of the $4D$ RPGM using the method of 
 duality and averaged Boltzmann factor.
 We have derived an equation to determine the location of the
 multicritical point (the accuracy threshold in the context of quantum
 error correction), which is expected to be exact (but is a conjecture,
 rigorously speaking).
 The structure of duality relations and the resulting equation for 
 the multicritical point coincide precisely with the corresponding 
 ones for the $2D$ RBIM. We could therefore generalize the
 generic duality relation originally proposed by Wegner \cite{Wegner} for non-random
 systems to the random case. In particular, the present
 argument can be extended to the model with $n$-dimensional
 objects in the $2n$-dimensional lattice. The same result  
 will be able to be derived for these models.

 We have elucidated the equivalence between the $4D$ RPGM 
 and the $4D$ plaquette qubit toric code 
 (and similarly for the $3D$ plaquette system).
 The phase boundary between the ordered and disordered
 phases on the Nishimori line gives the accuracy threshold of the toric
 code.
 From our study of the $4D$ RPGM, the accuracy threshold
 of the toric code with qubits on $4D$ plaquettes
 is expected to be given as $p=0.889972 \ldots$, the same result as 
 that of the $2D$ bond qubit system.

\begin{figure}
\begin{center}
  \includegraphics[width=15cm]{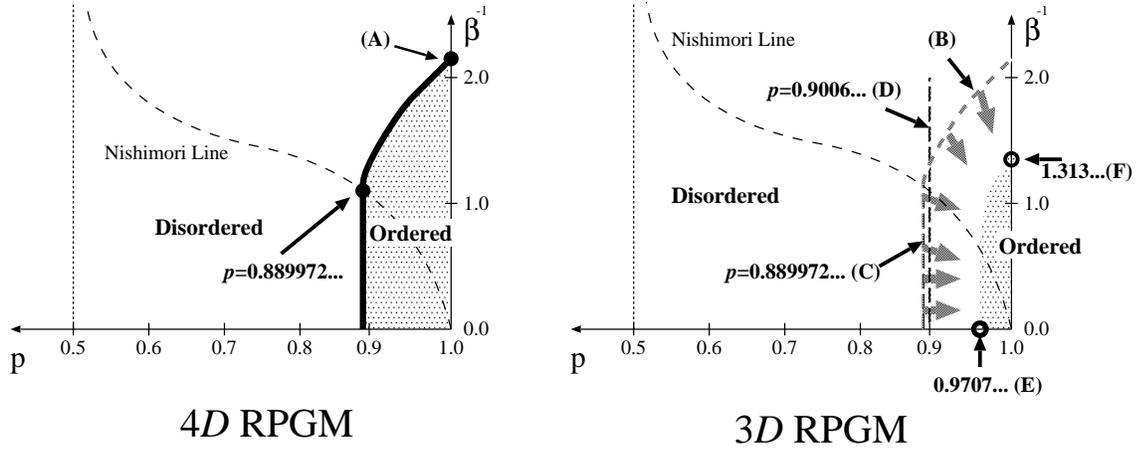}
\label{phasediagram2}
\caption{Phase diagram of the $4D$ and the $3D$ $\pm J$ RPGM.
Point (A): Critical point of non-random system obtained  exactly
 from the duality of non-random model.
 Line (B): Phase boundary of the $4D$ RPGM.
 Line (C): Lower bound of $p$ from the analysis of the $4D$ RPGM.
 Line (D): Lower bound of $p$ from the 
 analysis of the dual model of the $3D$ RPGM in Sec .6.
 Point (E): Critical point at $T=0$ by
 the numerical study in Ref. \cite{WHP}.
 Point (F): Critical point of the
 non-random model obtained by the numerical study of the $3D$ Ising model
 and by the duality between the $3D$ Ising model and the $3D$ plaquette
 gauge model.}
\end{center}
\end{figure}
 We may also expect that the phase boundary of the $4D$ RPGM 
 gives a limit to the boundary of the ordered phase for
 the $3D$ RPGM (Fig. 13) because the reduction
 of dimension always enhances disorder.
 Of course the phase boundary of the $4D$ RPGM gives the lower bound of
 probability of ferromagnetic phase for the $3D$ RPGM ($p=0.889972 \ldots$).
 The lower bound was also estimated from the $3D$ Ising model 
 using duality. The Griffiths inequality gives
 the lower bound $p=0.9006 \ldots$ on the assumption that the inequality
 holds for $n \rightarrow 0$ limit. This is a better bound 
 than the one from the study of the $4D$ RPGM ($p=0.889972 \ldots$).
 The transition point at $T=0$ (which is expected as the probability at
 the multicritical point) is estimated at $p=0.9707 \pm 0.0002$ 
 in Ref. \cite{WHP} by a numerical study,
 which is consistent with our inequality.

 As for the $4D$ plaquette toric code, 
 a {\it local} error correction procedure is also proposed
 in Ref. \cite{DKLP}.
 The method is to be contrasted with that in
 the present paper where we use {\it global} information to infer the
 error positions because one needs the information of
 the boundary of error sheet.
 The difference between these correction procedures is 
 as follows. Global procedure needs processes of fast classical computation
 to infer the real error sheet from its boundary, while
 the local procedure does not require such a classical
 computation because the local procedure needs only the local
 information of defects to remove all error sheets.
 On the other hand, the
 global procedure involves only a single observation
 to infer the error sheets and to correct errors.
 (Of course the observational errors occur in actual
 procedure and we must take them into consideration.)
 The local procedure requires many-step observations to
 erase all error sheets completely:
 The single correction step reduces the area of error sheet, but 
 not to wipe them out at once. At the same time, we must notice that 
 new error sheets may be generated during this process.
 If the rate of error generation is 
 larger than the error reduction rate of the correction procedure,
 the error correction will
 fail because homologically nontrivial error sheet will be formed
 by the accumulation of error sheets, which affects the encoded information.
 (Remember that the homologically nontrivial sheet is formed by real error sheets
 and correction sheets in the global procedure.) 
 Therefore the accuracy threshold also exists in the local procedure.
 In Ref. \cite{DKLP} an upper bound of 
 accuracy threshold of the error correction by the local procedure is estimated
 from the argument mentioned above, $p_c \simeq 1 - (4.8 \times 10 ^{-4})$.
 As pointed out in Ref. \cite{DKLP}, the local correction procedure cannot
 be applied to the bond-qubit system.
 We need at least four dimensions in order to
 perform the local correction procedure both
 for phase and bit-flip errors.

 \section*{Acknowledgement}
 The authors would like to thank Y. Ozeki, T. Sasamoto and T. Hamasaki
 for useful comments.
 This work was supported by the Grant-in-Aid for Scientific Research
 on Priority Area
 ``Statistical-Mechanical Approach to Probabilistic Information
 Processing''.


\begin{thebibliography}{99}
 \bibitem{lattice}  See e.g. C. Itzykson and J. -M. Drouffe,
         {\it Statistical Field Theory} (Cambridge University Press,
	Cambridge, 1989) and references therein.
 \bibitem{Kitaev} A. Yu. Kitaev, ``Quantum error correction with
	imperfect gates'' in {\it Proceedings of the Third
	International Conference on Quantum Communication and
	Measurement,} edited by O. Hirota, A. S. Holevo and C. M. Caves
	(Plenum, New York, 1997); A. Yu. Kitaev,
          Ann. Phys. {\bf 303} (2003) 2.
 \bibitem{DKLP} E. Dennis, A. Kitaev, A. Landahl and J. Preskill,
         J. Math. Phys. {\bf 43} (2002) 4452.
 \bibitem{WHP} C. Wang, J. Harrington and J. Preskill,
         Ann. Phys. {\bf 303} (2003) 31;\\
         Recently new results are obtained for 3D RPGM: G. Arakawa,
	 I. Ichinose, Ann. Phys. {\bf 311} (2004) 152;
         T. Ohno, G. Arakawa, I. Ichinose, T. Matsui, quant-ph/0401101. 
 \bibitem{NL} H. Nishimori, Prog. Theor. Phys. {\bf 66} (1981) 1169; 
         H. Nishimori, {\it Statistical Physics of Spin Glasses
	and Information Processing: An Introduction} (Oxford University
	Press, Oxford, 2001).
 \bibitem{Wegner} F. J. Wegner, J. Math. Phys. {\bf 12} (1971) 2259;
         R. Balian, J. M. Drouffe and C. Itzykson, Phys. Rev. {\bf D11}
         (1975) 2098;
         C. P. Korthals Altes, Nucl. Phys. {\bf B142} (1978) 315;
         A. Ukawa, P. Windey and A. H. Guth, Phys. Rev. {\bf D21} (1980) 1013.
 \bibitem{WuWang} F. Y. Wu and Y. K. Wang, J. Math. Phys. {\bf 17}
	  (1976) 439.
 \bibitem{MNN}
    J. -M. Maillard, K. Nemoto and H. Nishimori, J. Phys. {\bf A36}
	(2003) 9799; H. Nishimori and K. Nemoto, J. Phys. Soc. Jpn. {\bf 71}
         (2002) 1198.
 \bibitem{numerical} R. R. P. Singh and J. Adler,
	Phys. Rev. {\bf B54} (1996) 364;
	F. D. A. Aarao Reis, S. L. A. de Queiroz and R. R. dos Santos,
	Phys. Rev. {\bf B60} (1999) 6740;  A. Honecker, M. Picco and P. Pujol,
	Phys. Rev. Lett. {\bf 87} (2001) 047201; F. Merz and J. T. Chalker,
	Phys. Rev. {\bf B65} (2002) 054425;
         N. Ito and Y. Ozeki, Physica {\bf A321} (2003) 262;
         S. L. A. de Queiroz and R.
         Stinchcombe, Phys. Rev. {\bf B68} (2003) 144414.
 \bibitem{Griffiths} See e.g. R. B. Griffiths, ``Rigorous Results and
	Theorems'' in {\it Phase Transition and Critical Phenomena},
         Vol. 1, edited by C. Domb and M. S. Green (Academic Press,
	London, 1972).
 \bibitem{3DIsing}
 For resent results, see e.g. H. W. J. Bl\"ote, L. N. Shchur and
 A. L. Talapov, Int. J. Mod. Phys. {\bf C10} (1999) 1137;
 N. Ito, S. Fukushima, H. Watanabe and Y. Ozeki, ''Recent Development in
 Nonequilibrium Relaxation Method'' in
 {\it Computer Simulation Studies in Condensed-Matter Physics XIV,}
  edited by D. P. Landau, S. P. Lewis and H. -B. Schu\"uttler 
 (Springer-Verlag, Berlin, 2002);
 H. Arisue and T. Fujiwara, Nucl. Phys. {\bf B} (Proc. Suppl.) {\bf 119}
 (2003) 855;  Phys. Rev. {\bf E67} (2003) 066109.
 \end{thebibliography}
\end{document}